%% file: main.tex
\documentclass[twoside,journey]{IEEEtran}
\usepackage{makecell}
\usepackage{hyperref}
\usepackage{array}
\usepackage{graphicx,amssymb,amsmath}
\usepackage{multicol}
\usepackage[noadjust]{cite}
\usepackage{setspace}
\usepackage{subcaption}
\usepackage{graphicx}
\usepackage{float}
\usepackage {url}
\usepackage{stfloats}
\usepackage{amsthm,pifont}
\usepackage{flushend}
\usepackage{cases,subeqnarray}
\usepackage{bm,multirow,bigstrut}
\usepackage{amsmath, amsthm, amssymb}
\usepackage{textcomp}
\usepackage{latexsym,bm}
\usepackage{booktabs}
\usepackage{xcolor}
\usepackage{mathtools}
\usepackage{dsfont}
\usepackage{extarrows}
\usepackage{epsfig}
\usepackage{epsfig}
\usepackage{epstopdf}
\usepackage[noend]{algpseudocode}
\usepackage{algorithmicx,algorithm}
\usepackage{makecell}
\usepackage{colortbl}
\usepackage{booktabs}
\usepackage{graphicx} 
\usepackage{array}    
\usepackage{hyperref} 
\usepackage{booktabs}
\usepackage{enumitem}
\usepackage{tabularx}
\usepackage{booktabs}
\usepackage{diagbox}
\usepackage{tikz}
\usepackage{afterpage}
\usetikzlibrary{trees,shadows.blur}
\usepackage{forest}
\theoremstyle{plain}

\theoremstyle{plain}

\usepackage{amsmath}

\usepackage{rotating}

\newcommand{\ignore}[1]{{{\color{yellow} }}}
\renewcommand{\arraystretch}{1.5} 
\definecolor{blue-green}{rgb}{0.0, 0.87, 0.87}
\IEEEoverridecommandlockouts
\begin{document}

\title{Rethinking Wireless Communications through Formal Mathematical AI Reasoning\\}

\author{Changyuan Zhao, Jiacheng Wang, Dusit Niyato, Zan Li, Abbas Jamalipour, Shiwen Mao, \\Xianbin Wang, Dong In Kim
    \thanks{C. Zhao, J.~Wang, and D. Niyato are with the College of Computing and Data Science, Nanyang Technological University, Singapore (e-mail: zhao0441@e.ntu.edu.sg, jiacheng.wang@ntu.edu.sg, dniyato@ntu.edu.sg).}
            \thanks{Z. Li is with the State Key
Laboratory of Integrated Services Networks, Xidian University,
China (e-mail: zanli@xidian.edu.cn).}
    \thanks{A. Jamalipour is with the School of Electrical and Computer Engineering, University of Sydney, Australia, and with the Graduate School of Information Sciences, Tohoku University, Japan (e-mail: a.jamalipour@ieee.org).}
\thanks{S. Mao is with the Department of Electrical and Computer Engineering,
Auburn University, Auburn, USA (e-mail: smao@ieee.org).}
\thanks{X. Wang is with the Department of Electrical and Computer Engineering, Western University, Canada (e-mail: xianbin.wang@uwo.ca).}
\thanks{D. I. Kim is with the Department of Electrical and Computer Engineering, Sungkyunkwan University, South Korea (e-mail: dongin@skku.edu).}
\thanks{\textit{(Corresponding author: Dusit Niyato.)}}
}

\maketitle
\vspace{-1cm}

\begin{abstract}
Mathematical analysis has long underpinned wireless communication theory, yet the growing complexity of next-generation systems demands increasingly sophisticated reasoning from domain experts. Recent advances in AI mathematical reasoning, from formal theorem proving to large language model (LLM)-based derivation, offer a promising but largely unexplored path forward. Here we argue that wireless communications is a uniquely structured domain for formal AI reasoning, and propose a three-layer framework of verification, derivation, and discovery to rethink how wireless mathematical knowledge is established.
\end{abstract}


\IEEEpeerreviewmaketitle

\input{Introduction}
\input{section2}
\input{section3}

\input{section4}

\input{section5}

\bibliography{Ref}

\end{document}

%% file: Introduction.tex
\section{Introduction}



In 1948, Claude Shannon established the capacity of a noisy channel in a single landmark paper~\cite{shannon1948mathematical}, requiring little beyond probability theory and combinatorics. Nearly eight decades later, the mathematical foundations central to the design, evaluation and optimization of wireless systems have evolved profoundly. Characterizing the performance of massive multiple-input multiple-output (MIMO)~\cite{bjornson2017massive} calls for random matrix theory~\cite{couillet2011random}, analyzing coverage in heterogeneous networks~\cite{andrews2011tractable} relies on Laplace functionals over spatial point processes~\cite{haenggi2013stochastic}, and deriving the fundamental limits of integrated sensing and communication (ISAC)~\cite{liu2022integrated} involves Fisher information structures that couple waveform design, channel statistics, and estimation theory~\cite{liu2022survey}. 
Despite their critical importance in formulating diverse communication problems, solving them in next-generation wireless systems with growing complexity, dynamics and heterogeneity can be extremely challenging.

What unites these challenges is the depth and diversity of reasoning that they demand. A single derivation may draw on linear algebra, probability theory, stochastic geometry~\cite{haenggi2009stochastic, haenggi2013stochastic}, and non-convex optimization~\cite{luo2006introduction, liu2024survey} simultaneously, requiring sustained multi-step manipulation that must remain correct at every stage~\cite{tse2005fundamentals, cover1999elements}. As wireless systems continue to scale and become increasingly heterogeneous~\cite{bjornson2017massive, liu2022integrated}, the demand for sophisticated cross-domain reasoning is rapidly growing, often leading to complexities that cannot be handled human experts. 
This trend motivates the exploration of new tools and methodologies that can augment human expertise and accelerate theoretical progress.

\begin{figure*}[t]
  \centering
  \includegraphics[width= 0.8\linewidth]{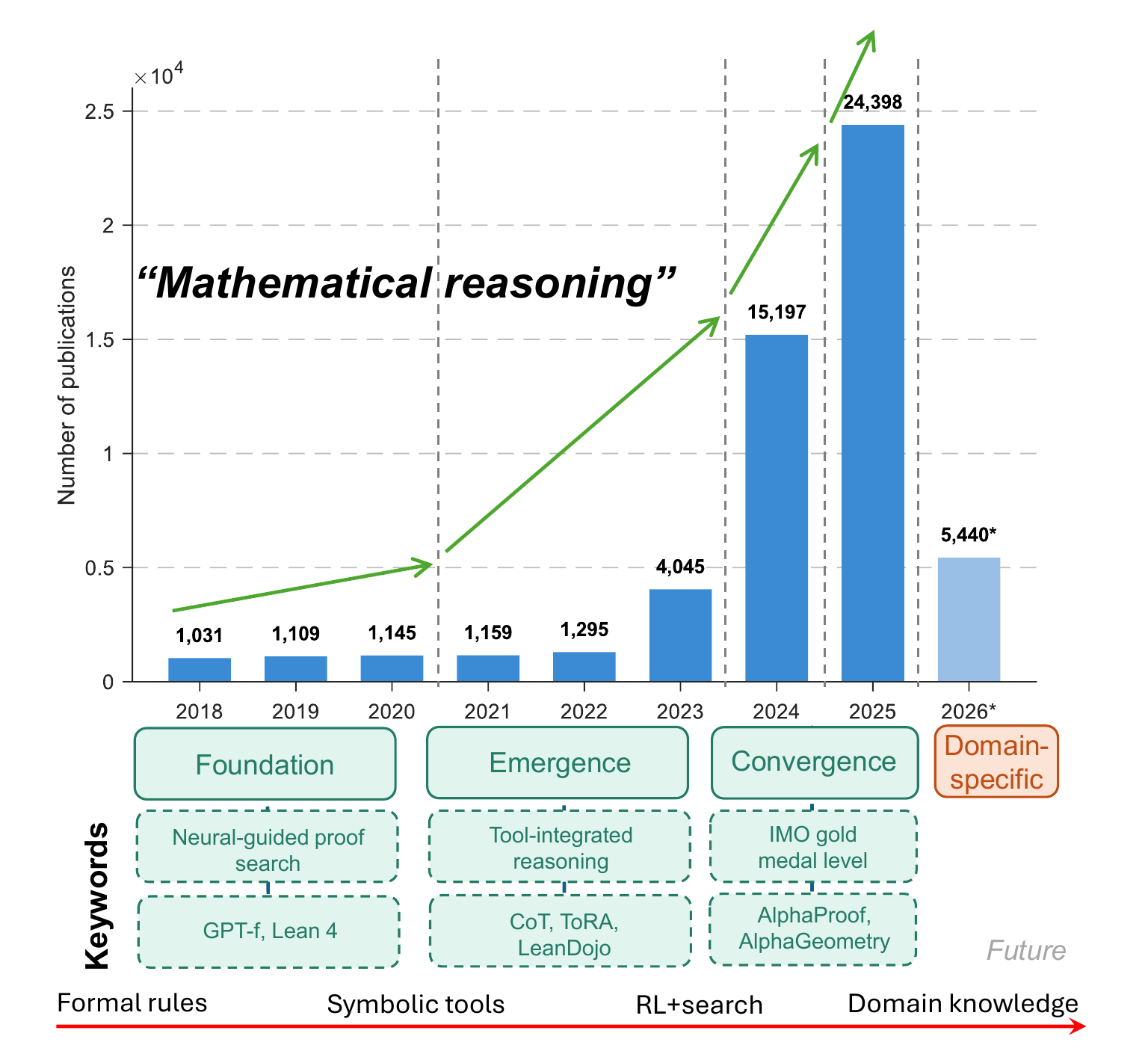}\\
  \caption{
  The rise of AI mathematical reasoning research. Annual number of publications indexed in Web of Science containing the keyword "mathematical reasoning" (2018–Apr. 10, 2026), highlighting four developmental phases. The Foundation phase (2018–2020) established formal rule-based systems and neural-guided proof search. The Emergence phase (2021–2022) introduced tool-integrated reasoning and structured prompting. The Convergence phase (2023–2025) marked the achievement of olympiad-level performance through reinforcement learning and search. A prospective Domain-specific phase envisions adapting these capabilities to structured scientific domains, such as wireless communications.}
  \label{fig:fig1}
\end{figure*}

In parallel, artificial intelligence (AI) is approaching an inflection point in mathematical reasoning. Formal theorem provers such as Lean~\cite{moura2021lean} and Coq~\cite{bertot2013interactive} now enable the construction and verification of multi-step proofs with machine-level precision, increasingly guided by neural proof search~\cite{lample2022hypertree,xin2024deepseek}. Neural-symbolic systems further extend this capability by combining learned pattern recognition with symbolic solvers to tackle problems in algebra, integration, and geometry~\cite{gou2023tora, unsal2024alphaintegrator, trinh2024solving}. Most notably, reinforcement learning-based systems have achieved silver-medal performance at the International Mathematical Olympiad through formally verified proofs~\cite{hubert2025olympiad}, while large language models (LLMs) trained on mathematical corpora demonstrate growing competence in multi-step derivation and quantitative reasoning~\cite{azerbayev2023llemma, shao2024deepseekmath, lewkowycz2022solving}. As shown in Fig.~\ref{fig:fig1}, the volume of publications on AI mathematical reasoning has grown rapidly since 2018. These advances signal a broader shift: AI is no longer confined to pattern recognition, but is emerging as a scalable engine for mathematical reasoning.

However, a notable gap persists. AI reasoning systems have been developed and benchmarked largely on competition mathematics and pure mathematical domains~\cite{hendrycks2021measuring, he2024olympiadbench, zheng2021minif2f}, while the structured, domain-specific problems of wireless communications remain largely unexplored. Wireless theory presents a distinct class of challenges. Its derivations require not only algebraic fluency, but also physically realistic assumptions, domain-specific approximations, and the integration of tools from estimation theory, stochastic processes, and optimization~\cite{tse2005fundamentals, li2025wirelessmathbench, liu2024survey}. At the same time, the wireless research community has yet to systematically engage with modern reasoning engines as analytical tools~\cite{he2024ai}. Bridging this gap could reshape the analytical workflow of wireless theory, enabling automated verification of classical results, accelerating multi-step derivations that currently demand extensive expert effort, and opening a path toward machine-assisted theoretical discovery.

Here, our perspective is that wireless communications provides a uniquely fertile ground for AI-assisted formal reasoning. Unlike competition mathematics, wireless problems are anchored in physical models with well-defined assumptions and measurable quantities, yet demand the same depth of multi-step logical reasoning~\cite{tse2005fundamentals, cover1999elements}. We organize this opportunity into three progressive roles for AI in wireless mathematics. Verification formalizes established results into machine-readable libraries. Derivation executes new analytical steps through symbolic and neural-symbolic reasoning. Discovery enables the generation and validation of new theoretical insights through autonomous conjecture. Through a case study on Cramér–Rao bound derivation in integrated sensing and communication, we offer early evidence that this vision is feasible with current reasoning systems, while identifying the key limitations that must be overcome. More broadly, this convergence between AI reasoning and wireless theory could open new possibilities for how wireless mathematical knowledge is constructed, verified, and extended.




%% file: section2.tex
\section{Mathematical Problems in Wireless Communications}

Wireless communication research encompasses a wide variety of 
mathematical problems involving multi-step analytical derivations, 
probabilistic reasoning, and complex optimization. Table~\ref{tab:math_problems} 
provides an overview of the principal problem classes introduced 
in this section, together with their key mathematical tools and 
opportunities for AI-assisted reasoning.

\begin{table*}[t]
\centering
\caption{Summary of mathematical problem classes in wireless communications, their analytical tools, and opportunities for AI-assisted reasoning.}
\label{tab:math_problems}
\renewcommand{\arraystretch}{1.4}
\begin{tabularx}{\textwidth}{p{2.8cm} p{4.0cm} p{4.2cm} p{5.2cm}}
\toprule
\textbf{Problem Class} & \textbf{Representative Problems} & \textbf{Key Mathematical Tools} & \textbf{Opportunities for AI-Assisted Reasoning} \\
\midrule
Linear Algebraic \& Statistical Inference
& \textbullet\ Signal detection~\cite{proakis2001digital, simon2004digital} \newline
  \textbullet\ Channel estimation~\cite{hassibi2003much, kay1993statistical} \newline
  \textbullet\ Beamforming design~\cite{bjornson2014optimal, shi2011iteratively}
& \textbullet\ Moment generating functions \newline
  \textbullet\ Matrix inversion lemmas \newline
  \textbullet\ Random matrix theory \newline
  \textbullet\ Semi-definite relaxation
& \textbullet\ Automated multi-fold integral evaluation \newline
  \textbullet\ Verification of asymptotic error characterizations \newline
  \textbullet\ Convergence proofs for iterative algorithms \\
\midrule
Information-Theoretic Analysis
& \textbullet\ Capacity derivation~\cite{telatar1999capacity, foschini1998limits} \newline
  \textbullet\ Coding bounds~\cite{polyanskiy2010channel} \newline
  \textbullet\ Secrecy capacity analysis~\cite{wyner1975wire, liu2010securing}
& \textbullet\ Entropy inequalities \newline
  \textbullet\ Jensen's inequality \newline
  \textbullet\ Saddle-point approximations \newline
  \textbullet\ Concentration inequalities
& \textbullet\ Automated bounding of mutual information \newline
  \textbullet\ Finite-blocklength derivation assistance \newline
  \textbullet\ Verification of multi-letter bounding arguments \\
\midrule
Stochastic Analysis \& Random Geometry
& \textbullet\ Outage probability~\cite{andrews2011tractable} \newline
  \textbullet\ Bit error rate analysis~\cite{simon2004digital, wang2003simple} \newline
  \textbullet\ Coverage and rate analysis~\cite{andrews2011tractable, dhillon2012modeling}
& \textbullet\ Stochastic geometry \newline
  \textbullet\ Laplace functionals \newline
  \textbullet\ Probability generating functionals \newline
  \textbullet\ Moment generating functions
& \textbullet\ Automated evaluation of Laplace functionals \newline
  \textbullet\ Closed-form simplification of infinite series \newline
  \textbullet\ Discovery of tractable approximations \\
\midrule
Large-Scale Optimization
& \textbullet\ Power allocation~\cite{yu2006dual, palomar2006tutorial} \newline
  \textbullet\ Scheduling~\cite{srikant2014communication} \newline
  \textbullet\ Interference management~\cite{cadambe2008interference, luo2008dynamic} \newline
  \textbullet\ Queue stability \& delay analysis~\cite{neely2010stochastic, wu2003effective}
& \textbullet\ Convex optimization \newline
  \textbullet\ Duality theory \newline
  \textbullet\ Successive convex approximation \newline
  \textbullet\ Game theory \newline
  \textbullet\ Lyapunov drift analysis \newline
  \textbullet\ Effective capacity theory
& \textbullet\ Automated duality gap characterization \newline
  \textbullet\ Verification of combinatorial relaxations \newline
  \textbullet\ Feasibility and degrees-of-freedom analysis \newline
  \textbullet\ Automated queue stability proofs \\
\bottomrule
\end{tabularx}
\end{table*}

\subsection{Linear Algebraic and Statistical Inference}
Signal processing in wireless communications gives rise to a 
class of structured inference problems used in linear algebra 
and statistical estimation~\cite{kay1993statistical, tse2005fundamentals}. 
Many of these problems originate from the canonical model 
$\mathbf{y} = \mathbf{H}\mathbf{x} + \mathbf{n}$, yet their 
analytical treatment involves intricate derivations whose 
complexity grows rapidly with system scale and channel 
heterogeneity.

Signal detection, for example, requires deriving closed-form 
or approximate error probability expressions by integrating 
over fading distributions such as Rayleigh, Rician, and 
Nakagami-$m$. These derivations often rely on moment 
generating function techniques and multi-fold integral 
evaluations~\cite{proakis2001digital, simon2004digital, wang2003simple}. 
A particularly influential result presented by Simon and Alouini 
establishes that the average bit error probability under 
generalized fading can be expressed in closed form via the 
moment generating function (MGF) evaluated at a fixed argument, reducing multi-fold 
integration over fading statistics to a single finite-range 
integral of the form 
$P_e = \frac{1}{\pi}\int_0^{\pi/2} M_\gamma\!\left(-\frac{1}{2\sin^2\theta}\right)d\theta$,
where $M_\gamma(\cdot)$ denotes the MGF of the received 
SNR~\cite{simon2004digital}. 
Channel estimation similarly demands careful manipulation of 
matrix inversion lemmas and Bayesian estimation frameworks, 
where the Woodbury matrix identity 
$({\mathbf{A}}+{\mathbf{U}}{\mathbf{C}}{\mathbf{V}})^{-1} = 
{\mathbf{A}}^{-1} - {\mathbf{A}}^{-1}{\mathbf{U}}
({\mathbf{C}}^{-1}+{\mathbf{V}}{\mathbf{A}}^{-1}{\mathbf{U}})^{-1}
{\mathbf{V}}{\mathbf{A}}^{-1}$
serves as a foundational lemma for deriving the MMSE estimation 
error covariance in closed form~\cite{kay1993statistical}. 
The analytical characterization of estimation error 
increasingly relies on asymptotic random matrix theory as 
antenna dimensions grow~\cite{hassibi2003much, bjornson2017massive, marzetta2010noncooperative}. 
Beamforming design further introduces non-convex optimization 
structures whose analytical treatment involves spectral 
decomposition, semi-definite relaxation, and convergence 
analysis of iterative algorithms such as weighted minimum mean 
square error (MMSE). A central result developed by Shi et al.\ 
establishes the equivalence between sum-rate maximization and 
a weighted MSE minimization problem, providing the analytical 
foundation for the convergence of the WMMSE 
algorithm~\cite{shi2011iteratively}.

Taken together, these problems require sustained multi-step 
algebraic reasoning across matrix analysis, probability theory, 
and optimization.


\subsection{Information-Theoretic Analysis}

Information theory establishes the fundamental limits of wireless 
communication systems, yet deriving these limits for practical 
channel models often requires substantial mathematical 
effort~\cite{shannon1948mathematical, cover1999elements}. The 
channel capacity is formally defined as $C = \max_{p(x)} I(X;Y)$, 
but evaluating this quantity rarely yields closed-form expressions 
without careful manipulation using tools from probability theory, 
entropy inequalities, and asymptotic 
analysis~\cite{el2011network}.

Capacity derivation often requires bounding mutual information 
under fading and interference, involving expectations over 
random channel matrices and the application of Jensen's 
inequality or saddle-point approximations. An important result 
by Telatar establishes that the ergodic capacity of an 
$n_t \times n_r$ MIMO channel with i.i.d.\ Rayleigh fading 
is given by
$C = \mathbb{E}\left[\log_2 \det\left(\mathbf{I}_{n_r} + 
\frac{\rho}{n_t}\mathbf{H}\mathbf{H}^H\right)\right]$,
a result whose derivation requires the joint eigenvalue density 
of Wishart matrices and careful application of Jensen's 
inequality to obtain computable capacity 
bounds~\cite{telatar1999capacity, foschini1998limits, goldsmith2002capacity}.
Coding bounds introduce further complexity by characterizing 
fundamental tradeoffs between rate, reliability, and blocklength. 
A central result presented by Polyanskiy, Poor, and Verd\'{u} 
establishes that in the finite-blocklength regime the maximum 
achievable rate satisfies
$R^*(n, \epsilon) \approx C - \sqrt{\frac{V}{n}}Q^{-1}(\epsilon) 
+ \mathcal{O}\!\left(\frac{\log n}{n}\right)$,
where $V$ is the channel dispersion and $Q^{-1}(\cdot)$ is the 
inverse Gaussian tail function, with the proof relying on 
Berry--Ess\'{e}en concentration inequalities applied to 
information density random 
variables~\cite{polyanskiy2010channel}. Secrecy capacity 
analysis adds another layer of difficulty, requiring auxiliary 
random variables and multi-letter bounding techniques; the 
foundational result by Wyner establishes that the secrecy 
capacity of the degraded wiretap channel equals 
$C_s = \max_{p(x)}\left[I(X;Y) - I(X;Z)\right]$,
a characterization whose proof demands careful reasoning over 
conditional entropy expressions and the construction of 
stochastic degradation 
arguments~\cite{wyner1975wire, liu2010securing}.

Across these problems, derivations typically involve chained 
applications of information-theoretic inequalities over multi-user 
and multi-antenna systems. As network size and channel 
heterogeneity increase, these analytical arguments grow 
progressively more intricate, creating a growing barrier to 
rigorous derivation and verification.

\subsection{Stochastic Analysis and Random Geometry}
Many wireless network performance analyses are based on stochastic 
models of spatially distributed nodes, drawing on tools from 
stochastic processes, random matrix theory, and stochastic 
geometry~\cite{haenggi2009stochastic, haenggi2013stochastic}. 
A common abstraction models network nodes as spatial point 
processes such as the homogeneous Poisson point process, 
enabling tractable yet realistic analysis of large-scale 
network behavior~\cite{baccelli2009stochastic}.

Outage probability analysis requires deriving the distribution 
of the received signal-to-interference-plus-noise ratio (SINR) over 
a random interference field, typically through Laplace 
functional techniques that condition on point process 
realizations and evaluate probability generating 
functionals~\cite{andrews2011tractable}. A landmark result 
developed by Andrews, Baccelli, and Ganti demonstrates that for a 
Poisson field of interferers with Rayleigh fading, the 
coverage probability admits the elegant closed-form expression
$P_c(\tau) = \pi\lambda \int_0^\infty 
\exp\!\left(-\pi\lambda r \tau^{2/\alpha} 
C(\alpha) - \mu \tau r^\alpha \sigma^2\right) dr$,
where $C(\alpha) = \frac{2\pi/\alpha}{\sin(2\pi/\alpha)}$ 
depends only on the path loss exponent $\alpha$, 
establishing stochastic geometry as a tractable framework 
for large-scale network analysis~\cite{andrews2011tractable}.
Bit error rate analysis introduces complementary challenges, 
requiring the averaging of conditional error probabilities 
over fading distributions through MGF methods and complex integral 
evaluation~\cite{simon2004digital}. 
Coverage and rate analysis further extends these derivations 
to multi-tier heterogeneous deployments, where a key result 
developed by Dhillon et al.\ shows that the aggregate interference 
from a $K$-tier heterogeneous network retains a tractable 
Laplace transform structure, allowing coverage and rate 
expressions to be decomposed tier-by-tier despite heterogeneous 
path loss exponents, fading parameters, and node 
densities~\cite{dhillon2012modeling}. These derivations 
lead to nested expectations and multi-dimensional integral 
expressions whose evaluation requires sustained algebraic 
reasoning over stochastic geometry and random process theory.

These analyses frequently yield results in the form of 
multi-fold integrals or infinite series whose closed-form 
simplification requires substantial algebraic manipulation.

\subsection{Large-Scale Optimization}
Wireless network design and resource management give rise to 
large-scale optimization problems that span convex and non-convex 
formulations as well as combinatorial structures~\cite{luo2006introduction, luo2008dynamic}. The general 
objective of maximizing sum-rate or energy efficiency subject to 
power, interference, and quality-of-service constraints requires 
tools from convex optimization, duality theory, and combinatorial 
optimization~\cite{palomar2006tutorial}.

Power allocation provides a canonical example: while the 
classical water-filling solution established by Cover and Thomas yields 
the closed-form optimal power allocation 
$p_k^* = \left(\mu - \frac{\sigma^2}{|h_k|^2}\right)^+$
across parallel channels, where $\mu$ is the water level 
determined by the total power constraint, multi-user scenarios 
require iterative methods whose convergence analysis involves 
non-trivial fixed-point arguments and duality gap 
characterization~\cite{yu2006dual}. Scheduling introduces 
combinatorial complexity, as the joint assignment of time, 
frequency, and spatial resources leads to problems whose 
relaxations rely on branch-and-bound procedures or carefully 
constructed greedy approximations with provable performance 
guarantees~\cite{srikant2014communication}. Interference 
management adds further analytical difficulty through the need 
to characterize feasibility conditions and degrees-of-freedom 
regions; a foundational result presented by Cadambe and Jafar 
establishes that the total degrees of freedom of the $K$-user 
interference channel equals $K/2$, achieved through interference 
alignment, a conclusion whose proof requires careful linear 
algebraic arguments over the space of beamforming 
vectors~\cite{cadambe2008interference}. The convergence 
properties of distributed algorithms such as successive convex 
approximation and game-theoretic update rules introduce further 
analytical challenges~\cite{luo2008dynamic, shi2011iteratively}. 
Queuing-theoretic formulations further enrich this landscape by 
introducing delay and stability constraints into resource 
allocation problems; a central result developed by Neely establishes 
that Lyapunov drift-plus-penalty algorithms achieve an 
$[\mathcal{O}(1/V), \mathcal{O}(V)]$ throughput-delay tradeoff, 
providing a rigorous analytical foundation for online resource 
allocation under time-varying wireless 
channels~\cite{neely2010stochastic, wu2003effective}.

These problems are often NP-hard in their original formulations, 
and deriving tight analytical characterizations of their solutions 
or performance bounds requires sustained reasoning across 
optimization theory, linear algebra, probability, and queuing 
theory.

\subsection{The Reasoning Bottleneck}
Despite this mathematical richness, the complexity of modern wireless systems has outpaced the capacity of traditional analytical workflows. Emerging technologies such as massive MIMO, ISAC, and large-scale heterogeneous networks~\cite{bjornson2017massive} require derivations that span multiple mathematical domains simultaneously, involving stochastic analysis, matrix theory, and non-convex optimization within a single framework. Theoretical results are predominantly obtained through manual derivation by domain experts, and formal verification of analytical conclusions remains uncommon in the wireless research community. As wireless systems continue to grow in scale and architectural complexity, this reliance on human-driven reasoning creates a widening gap between the mathematical demands of emerging systems and the tools available to analyze them.

\begin{figure*}[t]
  \centering
  \includegraphics[width= 0.8\linewidth]{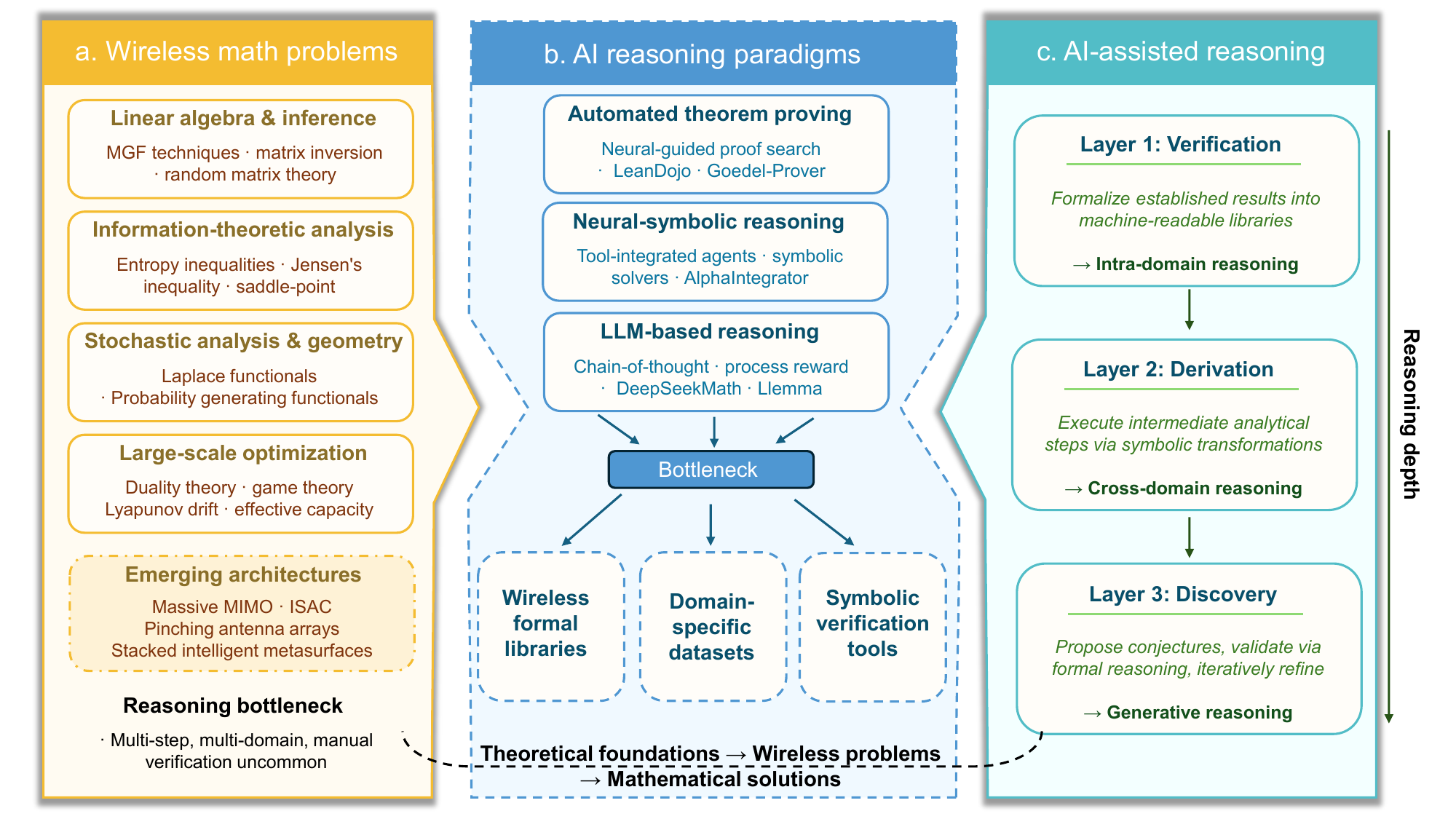}\\
  \caption{
From wireless mathematical problems to AI-assisted reasoning. (a) Wireless communication theory involves diverse mathematical problem classes, often requiring multi-step reasoning across domains. (b) Recent advances in AI reasoning span automated theorem proving, neural-symbolic reasoning, and LLM–based reasoning, yet these paradigms remain largely limited to general mathematical benchmarks and struggle with domain-specific structures in wireless theory. (c) Bridging this gap points toward a broader vision in which AI systems progressively support the verification of established results, the derivation of new analytical steps, and the discovery of theoretical insights in wireless mathematics.}
  \label{fig:fig2}
\end{figure*}

%% file: section3.tex
\section{Advances in AI Mathematical Reasoning}

Recent advances in AI have produced three broad paradigms for 
mathematical reasoning: automated theorem proving, neural-symbolic 
reasoning, and LLM-based reasoning~\cite{lu2023survey, yang2024formal, 
forootani2025survey}. Table~\ref{tab:ai_reasoning} summarizes these 
paradigms alongside their representative systems and opportunities 
for wireless mathematics.

\begin{table*}[t]
\centering
\caption{Summary of AI mathematical reasoning paradigms and their potential applications in wireless communications.}
\label{tab:ai_reasoning}
\renewcommand{\arraystretch}{1.4}
\begin{tabularx}{\textwidth}{p{2.8cm} p{3.2cm} p{4.2cm} p{4.8cm}}
\toprule
\textbf{Reasoning Paradigm} & \textbf{Representative Systems} & \textbf{Key Techniques} & \textbf{Opportunities for Wireless Mathematics} \\
\midrule
Automated Theorem Proving
& \textbullet\ Lean~\cite{moura2021lean}, Coq~\cite{bertot2013interactive} \newline
  \textbullet\ LeanDojo~\cite{yang2023leandojo} \newline
  \textbullet\ Goedel-Prover~\cite{lin2025goedel} \newline
  \textbullet\ Kimina-Prover~\cite{wang2025kimina}
& \textbullet\ Neural-guided proof search \newline
  \textbullet\ Reinforcement learning \newline
  \textbullet\ Retrieval-augmented proving \newline
  \textbullet\ Informal-to-formal pipelines
& \textbullet\ Verification of ergodic capacity expressions \newline
  \textbullet\ Automated proof of outage probability bounds \newline
  \textbullet\ Discovery of asymptotic scaling laws in massive MIMO \\
\midrule
Neural-Symbolic Reasoning
& \textbullet\ ToRA~\cite{gou2023tora} \newline
\textbullet\ MathSensei~\cite{das2024mathsensei} \newline
  \textbullet\ AlphaIntegrator~\cite{unsal2024alphaintegrator} \newline
  \textbullet\ AlphaGeometry~\cite{trinh2024solving} \newline
& \textbullet\ Tool-integrated reasoning \newline
  \textbullet\ Symbolic solver coupling \newline
  \textbullet\ Satisfiability-aided methods \newline
  \textbullet\ Neural conjecturing with symbolic deduction
& \textbullet\ Automated reformulation of non-convex power allocation \newline
  \textbullet\ Symbolic simplification of coverage probability expressions \newline
  \textbullet\ Closed-form approximation discovery for BER analysis \\
\midrule
LLM-based Reasoning
& \textbullet\ Llemma~\cite{azerbayev2023llemma}, Minerva~\cite{lewkowycz2022solving} \newline
  \textbullet\ DeepSeekMath~\cite{shao2024deepseekmath} \newline
  \textbullet\ MetaMath~\cite{yu2023metamath} \newline
  \textbullet\ Math-Shepherd~\cite{wang2024math}
& \textbullet\ Chain-of-thought prompting \newline
  \textbullet\ Process reward models \newline
  \textbullet\ Self-verification and correction \newline
  \textbullet\ Tree-of-thought search
& \textbullet\ Step-by-step outage probability derivations \newline
  \textbullet\ Automated ergodic capacity analysis \newline
  \textbullet\ Verified multi-step stochastic network modeling \\
\bottomrule
\end{tabularx}
\end{table*}

\subsection{Formal Languages and Verification Tools}

Formal reasoning over mathematical statements relies on tools that can
represent, manipulate, and verify logical arguments with machine-level
precision. Interactive proof assistants such as Lean~4, Coq, and
Isabelle/HOL provide formal logical frameworks in which mathematical
objects and their properties are encoded and mechanically verified,
ensuring correctness at every derivation step~\cite{moura2021lean, bertot2013interactive, nipkow2002isabelle}.
Built on these foundations, formal mathematical libraries such as
Mathlib accumulate large collections of verified lemmas across algebra,
analysis, and probability, providing reusable components for
formalization~\cite{baanen2025growing}.
Automated backends, including satisfiability modulo theories (SMT) solvers
and computer algebra systems, further extend these capabilities by
automatically solving algebraic and numerical subproblems.
Tools such as LeanDojo~\cite{yang2023leandojo} enable interaction between language
models and formal environments, allowing proposed proof steps to be
checked against the underlying logical system.
Together, these tools constitute the computational substrate for
AI-assisted mathematical reasoning.

\subsection{Automated Theorem Proving}

Formal proof assistants such as Lean and Coq enable machine-verifiable proofs by expressing mathematical statements and their derivations within a rigorous logical calculus~\cite{polu2020generative, jiang2021lisa}. Recent AI systems augment these frameworks with neural-guided proof search, where language models propose proof steps that are verified by the underlying formal system, substantially reducing the manual effort required for proof construction~\cite{lample2022hypertree, xin2024deepseek}. Reinforcement learning-based approaches further allow systems to iteratively conjecture and prove new results, with self-play mechanisms enabling models to autonomously generate and verify mathematical claims without human demonstration~\cite{dong2025stp, gloeckle2024abel}. Retrieval-augmented provers such as LeanDojo leverage proof search in curated lemma libraries, enabling compositional reasoning over previously established results~\cite{yang2023leandojo}, while growing library approaches such as LEGO-Prover progressively accumulate reusable proof modules to tackle increasingly complex theorems~\cite{wang2023lego}. Informal-to-formal pipelines, exemplified by Draft-Sketch-Prove, further improve automation by using language models to produce high-level proof sketches that guide formal verification systems~\cite{jiang2022draft}. Frontier models such as Goedel-Prover and Kimina-Prover, trained at scale with reinforcement learning, have pushed the state of the art on olympiad-level benchmarks, demonstrating that formal reasoning can extend beyond routine verification to genuine mathematical discovery~\cite{lin2025goedel, wang2025kimina, wei2024proving}. Existing theorem proving systems can already construct and verify formal proofs with machine-level precision. Extending these frameworks with wireless-specific libraries encoding stochastic channel models, expectation operators, and asymptotic approximations would enable automated verification and discovery of results such as ergodic capacity expressions, outage probability bounds, and asymptotic scaling laws in massive MIMO systems.

\subsection{Neural-symbolic Reasoning}

Neural-symbolic systems combine neural pattern recognition with symbolic manipulation to perform mathematical reasoning in a more flexible manner than purely formal approaches~\cite{karpas2022mrkl, olausson2023linc, qi2025large}. Tool-integrated reasoning agents such as ToRA and MathSensei couple LLMs with external symbolic solvers and program execution environments, allowing algebraic simplification and equation solving to be offloaded to reliable computational backends~\cite{gou2023tora, das2024mathsensei}. AlphaIntegrator extends this paradigm to symbolic integration, using transformer-based action search to construct step-by-step integration proofs that combine learned heuristics with formal verification~\cite{unsal2024alphaintegrator}. Program-of-thought and satisfiability-aided methods further translate reasoning tasks into executable programs or declarative logical constraints, improving reliability on multi-step numerical and combinatorial problems~\cite{gao2023pal,ye2023satlm}. Beyond algebraic domains, AlphaGeometry demonstrated that neural-symbolic systems can solve olympiad-level geometry problems without human demonstrations by interleaving neural conjecturing with symbolic deduction~\cite{trinh2024solving}. Collectively, these systems have demonstrated strong performance across algebra, geometry, and inequality reasoning benchmarks. 
Neural-symbolic systems have demonstrated strong performance in algebra, geometry, and inequality reasoning. Extending these pipelines to incorporate stochastic geometry and information-theoretic structures would enable automated reformulation of non-convex power allocation problems, symbolic simplification of coverage probability expressions, and closed-form approximation discovery for bit error rate analysis in complex fading environments.

\subsection{LLM-based Reasoning}

LLMs have demonstrated emerging capabilities in multi-step mathematical reasoning through chain-of-thought prompting, which encourages models to articulate intermediate derivation steps before producing a final answer~\cite{wei2022chain,sprague2024cot}. Numerical reasoning capabilities have been further strengthened through targeted pretraining strategies and skill injection techniques that endow models with stronger numeracy and arithmetic grounding~\cite{geva2020injecting, feng2021injecting}. Self-consistency and process reward models improve reliability by sampling multiple reasoning paths and selecting results that are consistent across attempts or verified at each intermediate step ~\cite{wang2022self, lightman2023let}, while self-verification and self-correction mechanisms allow models to autonomously detect and revise erroneous reasoning steps~\cite{zhang2025incentivizing,chen2025sets}. Test-time search strategies such as tree-of-thought enable deliberate exploration of branching solution paths, allowing models to backtrack and refine reasoning under uncertainty~\cite{yao2023tree}. To further scale mathematical training data, MetaMath bootstraps new question variants from existing problems, and Math-Shepherd provides dense step-level supervision without human annotation, both substantially improving reasoning robustness~\cite{yu2023metamath,wang2024math}. Domain-adapted models trained on large mathematical corpora, including Llemma, DeepSeekMath, and Minerva, have achieved strong results on formal reasoning, algebraic derivation, and quantitative problem solving~\cite{azerbayev2023llemma,shao2024deepseekmath,lewkowycz2022solving}. 
LLMs enhanced with chain-of-thought reasoning have shown emerging capability across multi-step mathematical derivations. Applying these systems to wireless mathematics could enable automated assistance in outage probability derivations, ergodic capacity analysis, and stochastic network modeling, with process reward mechanisms verifying each intermediate step to ensure derivation correctness across complex multi-antenna and heterogeneous network frameworks.

\subsection{Datasets}

Table~\ref{tab:benchmarks} summarizes the major benchmarks and corpora used to evaluate and train AI mathematical reasoning systems. A clear pattern emerges: existing resources concentrate almost entirely on competition and undergraduate-level pure mathematics. The sole wireless-oriented exception, WirelessMathBench~\cite{li2025wirelessmathbench}, focuses on mathematical modeling rather than multi-step derivations or formal verification, leaving the cross-domain reasoning chains that define wireless theory entirely uncovered.

\afterpage{%
\begin{sidewaystable*}[!ht]
\centering
\caption{Summary of mathematical reasoning benchmarks, datasets, and formal proof libraries.}
\label{tab:benchmarks}
\renewcommand{\arraystretch}{1.3}
\resizebox{\textheight}{!}{%
\begin{tabular}{llllll}
\toprule
\textbf{Dataset} & \textbf{Type} & \textbf{Size} & \textbf{Subject / Topic Distribution} & \textbf{Format} & \textbf{Link} \\
\midrule

\multicolumn{6}{l}{\textit{General Mathematical Reasoning}} \\[2pt]

MATH~\cite{hendrycks2021measuring}
  & Benchmark & 12,500 problems
  & \makecell[l]{\textit{Competition (AMC/AIME, levels 1--5)} \\ Prealgebra, Algebra, Intermediate Algebra,\\Precalculus, Geometry, Number Theory, Counting \& Probability}
  & \makecell[l]{\textbullet~Open-ended\\ \textbullet~Step-by-step solution}
  & \url{github.com/hendrycks/math} \\[4pt]\hline

GSM8K~\cite{cobbe2021training}
  & Benchmark & 8,500 problems
  & \makecell[l]{\textit{Middle school} \\ Elementary arithmetic word problems\\Addition, Subtraction, Multiplication, Division}
  & \makecell[l]{\textbullet~Open-ended\\ \textbullet~Natural language}
  & \url{github.com/openai/grade-school-math} \\[4pt]\hline

OlympiadBench~\cite{he2024olympiadbench}
  & Benchmark & 8,952 problems
  & \makecell[l]{\textit{Olympiad (IMO, IPhO, GaoKao)} \\ Mathematics 73\% (algebra, geometry, number theory), Physics 27\%}
  & \makecell[l]{\textbullet~Open-ended\\ \textbullet~Expert-annotated solution}
  & \url{github.com/OpenBMB/OlympiadBench} \\[6pt]\hline\hline

\multicolumn{6}{l}{\textit{Formal Theorem Proving Benchmarks}} \\[2pt]

MiniF2F~\cite{zheng2021minif2f}
  & Benchmark & 488 problems
  & \makecell[l]{\textit{High school -- Olympiad} \\ Algebra, number theory, combinatorics}
  & \makecell[l]{\textbullet~Formal proof\\~~(Lean / Isabelle / Metamath)}
  & \url{github.com/openai/miniF2F} \\[4pt]\hline

FormalMATH~\cite{yu2025formalmath}
  & Benchmark & 5,560 problems
  & \makecell[l]{\textit{High school -- Olympiad} \\ Algebra, geometry, calculus, number theory, discrete math}
  & \makecell[l]{\textbullet~Formal proof (Lean~4)}
  & \url{arxiv.org/abs/2505.02735} \\[6pt]\hline\hline

\multicolumn{6}{l}{\textit{Formal Proof Libraries and Corpora}} \\[2pt]

NaturalProofs~\cite{welleck2021naturalproofs}
  & Corpus & \makecell[l]{$\sim$32,000 theorems \\ 15,000 definitions}
  & \makecell[l]{\textit{Undergraduate -- research} \\ Broad pure mathematics (ProofWiki, Stacks project, textbooks)\\Analysis, Algebra, Number theory, Topology}
  & \makecell[l]{\textbullet~Natural language proof \\ \textbullet~References}
  & \url{github.com/wellecks/naturalproofs} \\[4pt]\hline

LeanDojo~\cite{yang2023leandojo}
  & Library / Corpus & \makecell[l]{98,734 theorems (Lean~3) \\ 122,517 theorems (Lean~4)}
  & \makecell[l]{\textit{Undergraduate -- research} \\ Full Mathlib coverage: Algebra, Analysis, \\Topology, Number theory, Combinatorics}
  & \makecell[l]{\textbullet~Formal proof (Lean)\\ \textbullet~Tactic-level annotation}
  & \url{github.com/lean-dojo/LeanDojo} \\[4pt]\hline

IsarStep~\cite{li2020isarstep}
  & Corpus & 204,000 lemmas
  & \makecell[l]{\textit{Undergraduate -- research} \\ Logic, analysis, algebra, cryptography, CS (Isabelle Archive of Formal Proofs)}
  & \makecell[l]{\textbullet~Isabelle/HOL proof steps}
  & \url{github.com/Wenda302/IsarStep} \\[4pt]\hline

Lean Workbook~\cite{ying2024lean}
  & Corpus & \makecell[l]{57,000 problems \\ (5,000 with proofs)}
  & \makecell[l]{\textit{High school -- Olympiad} \\ Competition math (AOPS, IMO): Algebra, Inequalities, Number theory}
  & \makecell[l]{\textbullet~Formal statement (Lean~4)\\ \textbullet~Autoformalized}
  & \url{github.com/InternLM/InternLM-Math} \\[4pt]\hline

ProofNet~\cite{azerbayev2023proofnet}
  & Benchmark & 371 problems
  & \makecell[l]{\textit{Undergraduate} \\ Real and complex analysis, linear algebra, abstract algebra, topology}
  & \makecell[l]{\textbullet~Formal statement (Lean~3) \\ \textbullet~Natural language proof}
  & \url{github.com/zhangir-azerbayev/ProofNet} \\[4pt]\hline

LEAN-GitHub~\cite{wu2024lean}
  & Corpus & \makecell[l]{28,597 theorems \\ 218,866 tactics}
  & \makecell[l]{\textit{High school -- research} \\ Diverse GitHub repos: Logic, Algebra, Matroid theory,\\Data structures, Olympiad problems}
  & \makecell[l]{\textbullet~Formal proof (Lean~4) \\ \textbullet~Tactic steps}
  & \url{github.com/InternLM/InternLM-Math} \\[6pt]\hline\hline

\multicolumn{6}{l}{\textit{Wireless Communications}} \\[2pt]

WirelessMathBench~\cite{li2025wirelessmathbench}
  & Benchmark & 587 problems
  & \makecell[l]{\textit{Research} \\ System models: RIS (48\%), MIMO (30\%), UAV/ISAC (15\% each), others; \\ Problem types: Beamforming (45\%), Channel Estimation (30\%), \\Performance Analysis (20\%), others}
  & \makecell[l]{\textbullet~Masked fill-in-the-blank\\ \textbullet~Full equation completion}
  & \url{lixin.ai/WirelessMathBench} \\

\bottomrule
\end{tabular}%
}
\end{sidewaystable*}%
}


\subsection{The Opportunity for AI-Assisted Wireless Mathematics}


Across these paradigms, AI reasoning systems have been developed and evaluated almost exclusively on general mathematical benchmarks. The distinct mathematical structures of wireless communications, including stochastic channel modeling, random matrix theory, stochastic geometry, and information-theoretic analysis, remain largely untapped. Closing this gap requires three foundations: formalized wireless mathematical libraries, domain-specific reasoning datasets, and standardized evaluation benchmarks. Together, they would enable AI systems to verify, derive, and discover theoretical results across next-generation wireless networks. Fig.~\ref{fig:fig2} provides a unified illustration of this convergence, mapping the principal classes of wireless mathematical problems to the three AI reasoning paradigms and the proposed three-layer framework for AI-assisted wireless mathematics.

%% file: section4.tex
\section{AI-Assisted Reasoning for Wireless Mathematics}

\subsection{The Difficulty Spectrum of Wireless Mathematical Problems}

 

Wireless communication theory encompasses a spectrum of mathematical challenges of increasing depth and complexity. At the most accessible level, problems involve formalizing system descriptions into precise mathematical abstractions, such as casting resource allocation tasks as constrained optimization programs or representing channel dynamics through stochastic models~\cite{luo2006introduction, palomar2006tutorial}. Beyond formulation, a more demanding class arises in analytical derivation, where multi-fold integrals over fading distributions, Laplace functionals in stochastic geometry, or asymptotic approximations of capacity must be carefully evaluated and simplified~\cite{simon2004digital,andrews2011tractable}. At the highest level lies mathematical proof, requiring rigorous logical arguments to establish fundamental properties, such as the convergence of iterative algorithms, the degrees of freedom of interference channels, or finite-blocklength performance limits~\cite{cadambe2008interference, polyanskiy2010channel}.

These layers reflect increasing technical difficulty and a shift in reasoning, from model formulation to analytical manipulation and formal proof. As wireless systems scale, this progression exposes a growing bottleneck in human-driven analysis. We thus organize AI-assisted wireless reasoning into three roles: verification, derivation, and discovery.

\subsection{Layer 1: AI-Assisted Verification of Existing Results}

Recent advances in formal reasoning systems have made automated proof verification increasingly practical. Systems such as LeanDojo~\cite{yang2023leandojo} and DeepSeek-Prover~\cite{xin2024deepseek} can construct and check multi-step proofs by leveraging large-scale formal libraries and learned proof search.

This capability is directly relevant to wireless theory, where many classical closed-form expressions and theoretical bounds already constitute the mathematical foundation of the field. For instance, expressions such as the ergodic capacity of MIMO channels~\cite{telatar1999capacity} or coverage probability in stochastic geometry~\cite{andrews2011tractable} are built on standard transformations involving expectations, inequalities, and asymptotic approximations. Similar formalization efforts have begun in adjacent scientific domains: recent work has developed Lean-based libraries for International System of Units unit systems and dimensional analysis in physics~\cite{li2025lean4physics}, suggesting that domain-specific formal libraries are feasible beyond pure mathematics. The first layer, therefore, is to formalize these established results into machine-readable libraries, such as Lean-based repositories, so that existing reasoning tools and AI models can directly read and reuse them. This layer operates primarily in an intra-domain setting, where reasoning proceeds by invoking existing wireless results.

\subsection{Layer 2: AI-Assisted Derivation of New Analytical Steps}

A central bottleneck in wireless analysis lies in executing intermediate analytical steps in addition to formulating the problem itself. Neural-symbolic systems address this bottleneck by equipping reasoning agents with symbolic tools such as Mathematica~\cite{wolfram2003mathematica}, SMT solvers~\cite{de2008z3}, and formal proof checkers~\cite{moura2021lean}, which execute intermediate analytical steps from established results through rigorous symbolic inference. For instance, AlphaIntegrator~\cite{unsal2024alphaintegrator} constructs step-by-step integration proofs by chaining known symbolic rules, while tool-integrated agents such as ToRA~\cite{gou2023tora} invoke external solvers to verify and extend intermediate results during multi-step reasoning.

Wireless derivations often rely on results beyond the domain itself, drawing on tools from stochastic geometry, matrix analysis, random matrix, and optimization~\cite{couillet2011random}. In this layer, AI-based reasoning systems operate in a problem-driven manner: given a new analytical task, they identify appropriate mathematical tools or transfer relevant results from existing domains, and integrate them into the derivation process through symbolic transformations and composition. This recasts derivation as cross-domain reasoning, where wireless problems are solved by adaptively leveraging knowledge across multiple mathematical domains.

\subsection{Layer 3: Discovery of New Theoretical Results}
 
The most forward-looking direction is the autonomous discovery of new theoretical results in wireless. Reinforcement learning-based provers, such as the Goedel-Prover~\cite{lin2025goedel}, have demonstrated the ability to generate and prove non-trivial mathematical statements without human supervision.

Wireless communication provides a structured domain where such capabilities may be particularly impactful. Many open problems, such as tighter finite-blocklength bounds~\cite{polyanskiy2010channel} or refined degrees-of-freedom characterizations~\cite{cadambe2008interference}, are defined over well-specified mathematical models with clear constraints. In this layer, AI-based reasoning systems move beyond reusing or building upon existing results, and instead operate through a generative process: proposing new conjectures, validating them via formal or symbolic reasoning, and iteratively refining them into provable statements. This establishes a novel regime of reasoning in which new theoretical results are discovered rather than derived from existing knowledge. As emerging architectures such as pinching antenna arrays and stacked intelligent metasurfaces (SIMs) give rise to new analytical frameworks, AI-assisted reasoning may help derive the underlying theoretical foundations before they are fully established by conventional analysis.

\subsection{Lesson Learned}

Collectively, the three layers suggest not merely a redistribution of analytical labor but a redefinition of how wireless mathematical knowledge is produced. As AI systems begin to participate in verification, derivation, and ultimately discovery, the role of human researchers shifts from executing analytical steps to shaping the structure and interpretation of reasoning itself. Unlike pure mathematics, wireless reasoning must operate over signal models, channel statistics, and physical assumptions that cannot be discarded without losing the meaning of the result. Progress, therefore, hinges less on general-purpose reasoning performance than on the development of wireless-specific foundations, including formal libraries, domain-grounded datasets, and reasoning frameworks capable of sustaining physically interpretable derivations at every layer.

\begin{figure*}[t]
  \centering
  \includegraphics[width= 0.95\linewidth]{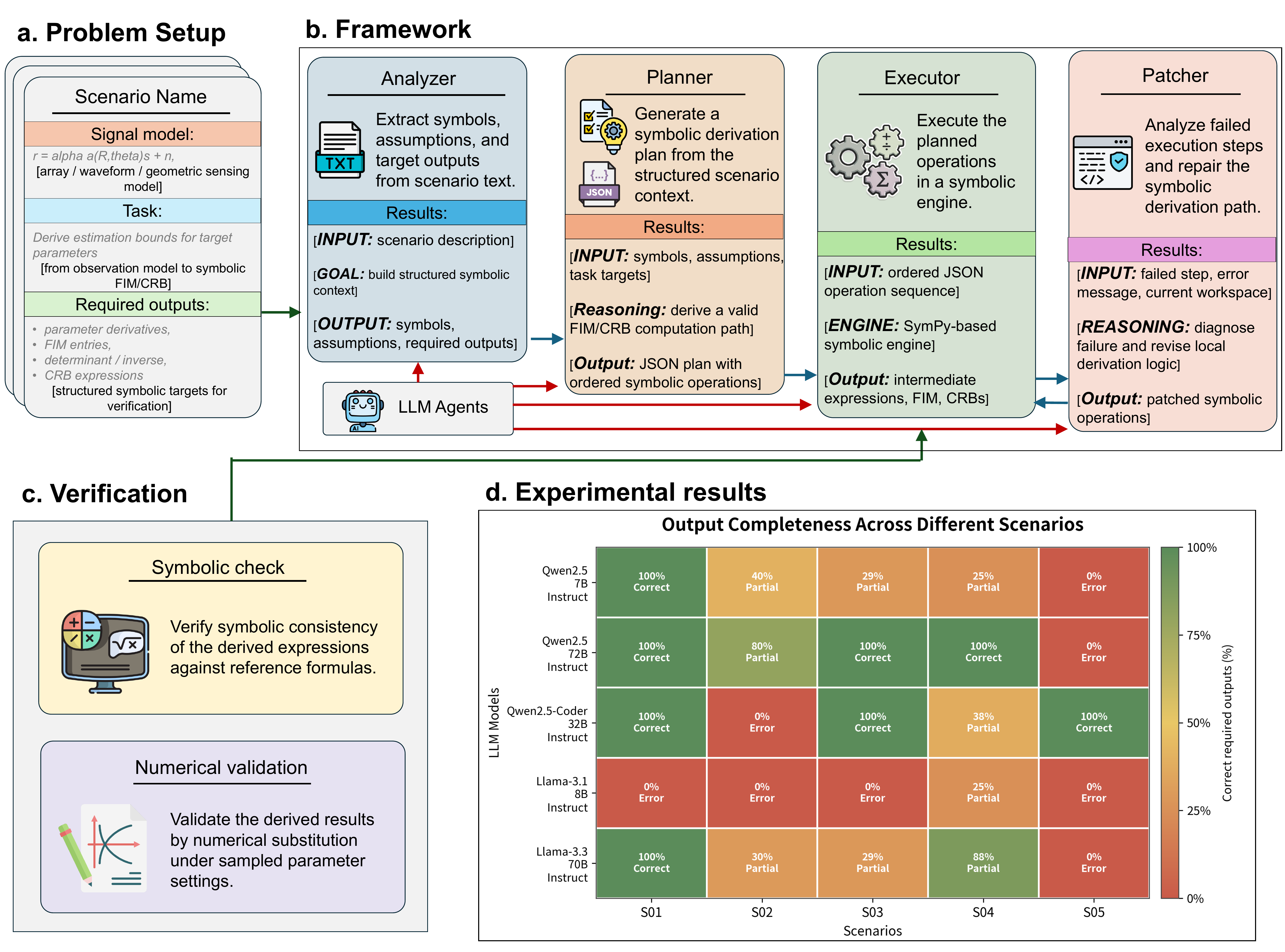}\\
  \caption{
A role-based LLM reasoning framework for Cramér–Rao bound derivation in ISAC. (a) Each problem is specified by a signal model, target parameters, and required symbolic outputs. (b) The framework decomposes the derivation into four sequential roles: the Analyzer extracts symbols and assumptions; the Planner generates an ordered sequence of symbolic operations; the Executor carries out these operations in a SymPy-based engine; and the Patcher diagnoses and repairs failed steps. (c) Derived results are validated through symbolic verification against reference formulas and numerical validation under sampled parameter settings. (d) LLM reasoning accuracy across five ISAC scenarios (S01–S05) for five language models of varying scale.}
  \label{fig:fig3}
\end{figure*}

\section{Case Study: Role-Based AI Reasoning for Mathematical Analysis in ISAC}

As a representative case study, we consider the Cramér–Rao bound (CRB) derivation in ISAC~\cite{liu2022integrated}. CRB analysis in ISAC constitutes a structured analytical problem that requires the coordinated use of signal modeling, parameter differentiation, Fisher information matrix construction, matrix inversion, and asymptotic or closed-form simplification~\cite{liu2022survey}. Its derivation is highly sensitive to intermediate correctness, since even minor inconsistencies in derivatives, coefficients, or summation terms can propagate to the final bound and alter its physical interpretation.

To support this task, we develop a staged LLM reasoning framework in which different components play complementary roles within the derivation pipeline, as illustrated in Fig.~\ref{fig:fig3}.
\begin{itemize}
    \item \textbf{Analyzer} parses the scenario description, extracts symbols and helper quantities, and seeds the symbolic workspace before derivation begins.
    \item \textbf{Planner} produces a full derivation plan in the form of executable symbolic operations for the target CRB problem.
    \item \textbf{Executor} carries out the planned operations locally and step by step using symbolic tools such as SymPy, while recording intermediate results and failures.
    \item \textbf{Patcher} analyzes failed execution steps, checks for any errors arising from the derivation, and then proposes corrected continuation steps.
\end{itemize}

We instantiate the framework on a suite of 5 representative CRB problems in ISAC and radar sensing, including near-field uniform linear array (ULA) angle-range estimation, frequency diverse array (FDA)-based range-angle coupling, clutter-limited CRB analysis, three-dimensional near-field localization, and joint velocity-acceleration estimation\footnote{Specific models and full implementation available at \url{https://github.com/ChangyuanZhao/Wireless_mathematical_reasoning}.}.
These scenarios were selected to cover a diverse set of analytical regimes, ranging from scalar and vector parameter estimation to coupled Fisher information structures and matrix-valued inversions~\cite{kay1993statistical}. Each scenario is specified in a structured form comprising the signal model, the unknown parameters of interest, and the analytical tool available for derivation, such as the Slepian–Bangs formula for complex Gaussian observations.
To make the mathematical structure of the case study explicit, Table~\ref{tab:nearfield_ula_case} summarizes the key formulations used in the representative near-field ULA CRB derivation.

\begin{table*}[t]
\centering
\caption{Representative near-field ULA CRB case study used in our benchmark. We distinguish between the information provided in the scenario specification and the symbolic quantities that the LLM is required to derive.}
\label{tab:nearfield_ula_case}
\renewcommand{\arraystretch}{1.15}
\begin{tabular}{p{2.8cm} p{3.2cm} p{10.0cm}}
\toprule
\textbf{Section} & \textbf{Item} & \textbf{Content} \\
\midrule

\multirow{4}{*}{Given in scenario}
& Scenario
& Near-field ULA joint angle--range estimation under a second-order spherical wavefront model. \\

& Steering vector
& $\displaystyle
[a(R,\theta)]_m
=
\exp\!\left(
j\frac{2\pi}{\lambda}
\left[
md\sin\theta-\frac{(md)^2\cos^2\theta}{2R}
\right]
\right),\quad m=0,\ldots,M-1
$ \\

& Observation model
& $\displaystyle
\mathbf{r}=\alpha\,\mathbf{a}(R,\theta)\,s+\mathbf{n},
\qquad
\mathbf{n}\sim\mathcal{CN}(\mathbf{0},\sigma^2\mathbf{I}_M)
$ \\

& Fisher information formula
& $\displaystyle
F_{ij}
=
\frac{2|\alpha|^2}{\sigma^2}\,
\mathrm{Re}\!\left[
\left(\frac{\partial \mathbf{a}}{\partial \eta_i}\right)^H
\left(\frac{\partial \mathbf{a}}{\partial \eta_j}\right)
\right],
\qquad
\boldsymbol{\eta}=[\theta,R]^T
$ \\

\midrule

\multirow{8}{*}{Target outputs}
& $d\_\phi\_m\_d\_\theta$
& Derivative of the steering phase with respect to $\theta$. \\

& $d\_\phi\_m\_d\_R$
& Derivative of the steering phase with respect to $R$. \\

& $F_{\theta\theta}$
& Fisher information entry for angle. \\

& $F_{RR}$
& Fisher information entry for range. \\

& $F_{\theta R}$
& Cross term of the Fisher information matrix. \\

& $\det(\mathbf{F})$
& Determinant of the $2\times2$ Fisher information matrix. \\

& $\mathrm{CRB}_{\theta}$
& Cramér--Rao bound for angle estimation. \\

& $\mathrm{CRB}_{R}$
& Cramér--Rao bound for range estimation. \\

\midrule

\multirow{7}{*}{Symbol definitions}
& $M$
& Number of antennas in the uniform linear array. \\

& $d$
& Inter-element spacing. \\

& $\lambda$
& Signal wavelength. \\

& $\theta$
& Target angle. \\

& $R$
& Target range. \\

& $\alpha$
& Complex path gain, with $|\alpha|=1$. \\

& $\sigma^2$
& Noise variance. \\

\bottomrule
\end{tabular}
\end{table*}

\begin{figure}[t]
  \centering
  \includegraphics[width= 0.95\linewidth]{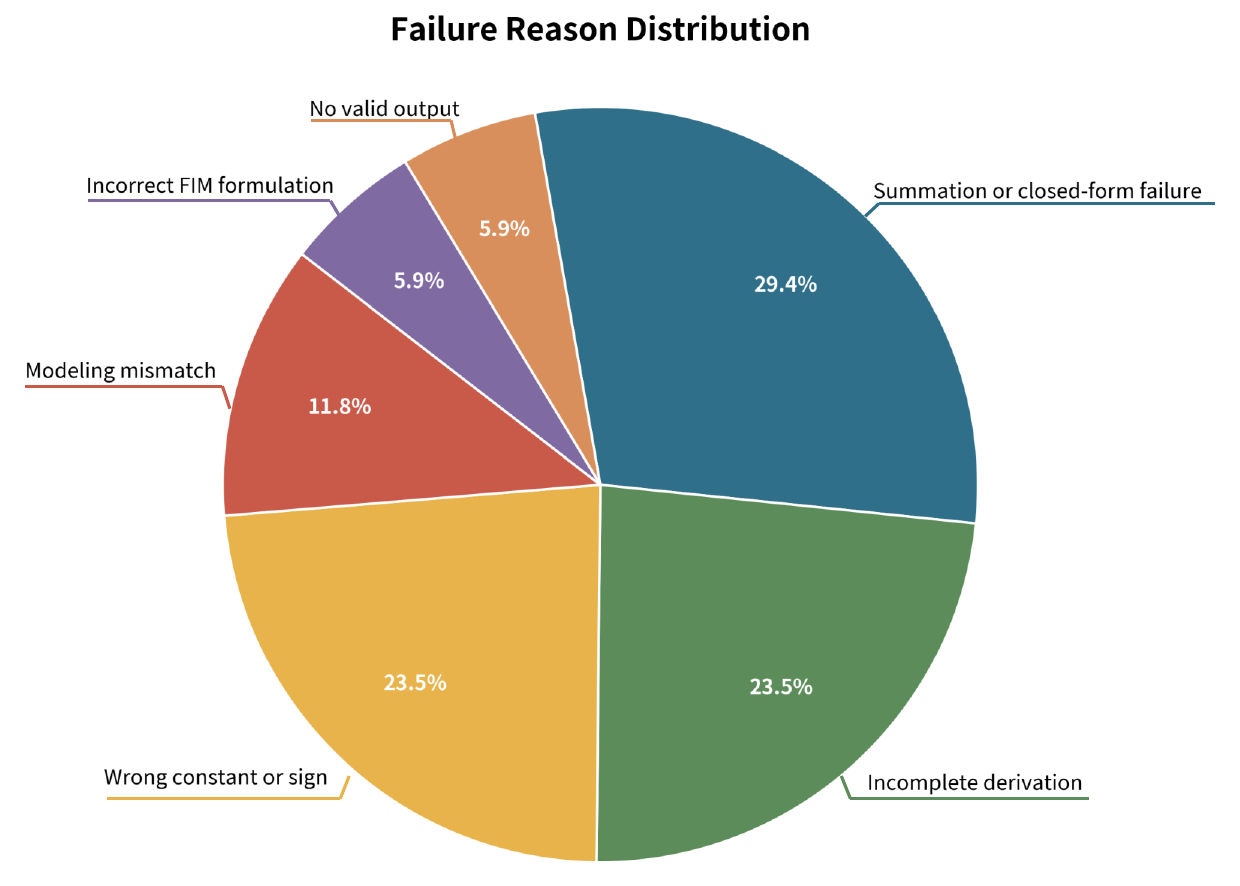}\\
  \caption{
  Distribution of failure modes in LLM-assisted CRB derivation.
Analysis of derivation failures across all model–scenario combinations, categorized by root cause. The most frequent failure modes are summation or closed-form errors (29.4\%) and incomplete derivations (23.5\%), followed by incorrect constants or signs (23.5\%), modeling mismatches (11.8\%), incorrect FIM formulation (5.9\%), and cases producing no valid output (5.9\%). These results highlight that the dominant sources of error lie in algebraic manipulation and derivation completeness rather than in problem formulation, suggesting that stronger symbolic execution and step-level verification are the most impactful targets for improvement.}
  \label{fig:fig4}
\end{figure}

\begin{figure}[t]
  \centering
  \includegraphics[width= 0.95\linewidth]{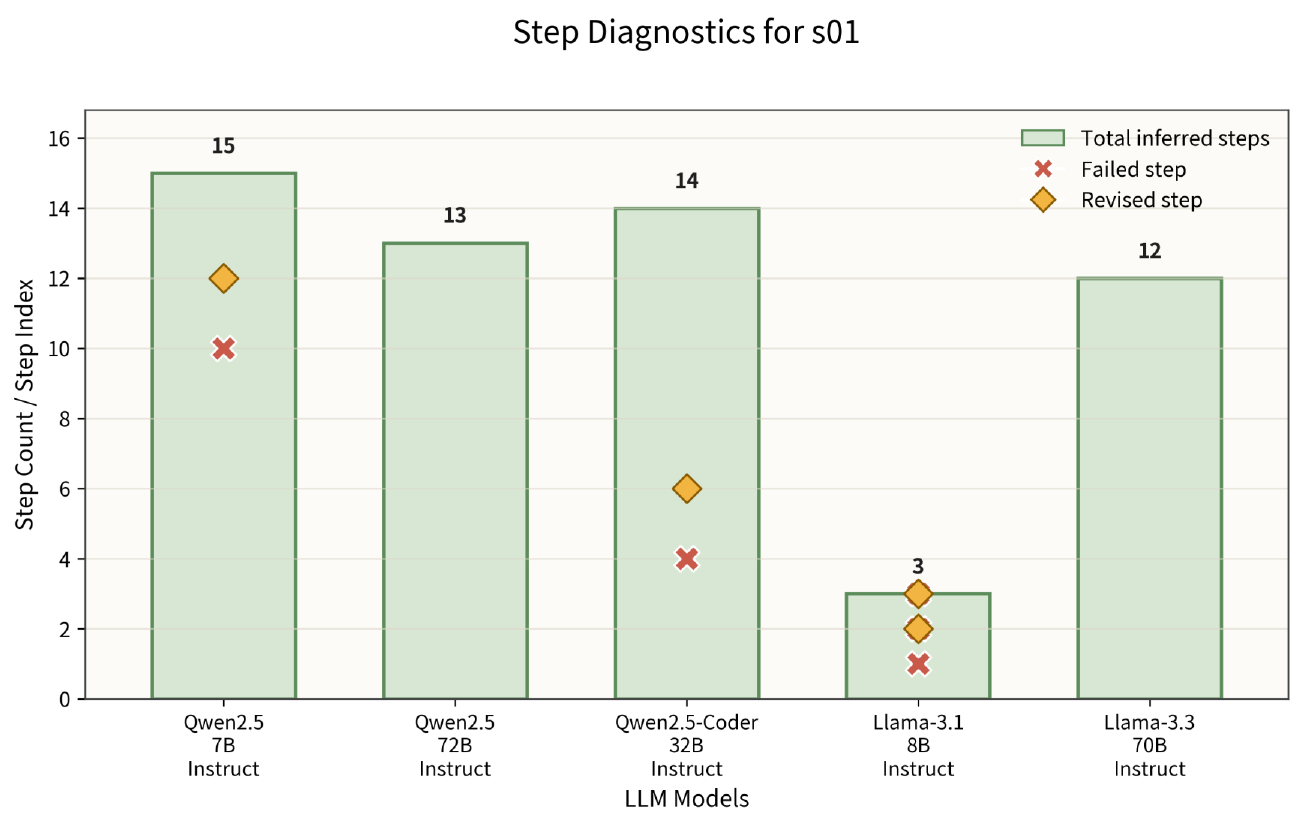}\\
  \caption{
  Step-level diagnostics of LLM reasoning for scenario S01.
For each of the five language models evaluated on the near-field ULA angle-range estimation scenario (S01), the chart shows the total number of inferred derivation steps alongside the indices at which failures and subsequent revisions occurred. Models with larger parameter counts tend to generate longer derivation chains and encounter failures at later stages, indicating deeper engagement with the problem structure. The Patcher agent's revision steps (diamonds) demonstrate the framework's capacity for self-repair, with successful patches enabling continued derivation beyond initial failure points.}
  \label{fig:fig5}
\end{figure}

The evaluation is structured along two complementary axes, with results summarized in Figs. ~\ref{fig:fig3}(d),~\ref{fig:fig4}, and~\ref{fig:fig5}. The first examines reasoning performance across five language models spanning a range of scales. As illustrated in Fig. 3(d), larger models and those with stronger coding proficiency exhibit more consistent derivation accuracy. A closer analysis of failure modes in Fig.~\ref{fig:fig4} reveals that the errors are predominantly algebraic in nature. In contrast, errors arising from problem formulation are comparatively limited. This distribution indicates that the principal bottleneck lies not in understanding the task, but in reliably executing multi-step symbolic manipulations, highlighting symbolic execution as a critical target for improvement.

The second axis investigates reasoning depth through step-level diagnostics, as shown in Fig.~\ref{fig:fig5}. Larger models tend to produce longer and more structured derivation trajectories, suggesting a deeper engagement with the underlying analytical process. Importantly, the introduction of the Patcher agent enables effective recovery from intermediate failures, allowing derivations to progress beyond initial breakdown points. This behavior underscores the importance of iterative refinement mechanisms for extending reasoning beyond single-pass generation.

Collectively, these observations suggest that enhancing symbolic backends and introducing fine-grained, step-level verification could substantially improve the reliability of AI-assisted mathematical derivation. More broadly, such reasoning frameworks are not limited to CRB analysis, but could be extended to a wider class of wireless mathematical problems, including information-theoretic analysis and stochastic geometry. Realizing this potential will require tighter integration between learning-based reasoning models and domain-specific symbolic tools, alongside the development of structured benchmarks tailored to wireless theory.

%% file: section5.tex
\section{Challenges and Research Opportunities}

Despite its promise, AI-assisted reasoning for wireless communications faces several challenges that also define key research opportunities.

\begin{itemize}
\item Formalization of wireless mathematical problems. Fundamental mathematical formulations in wireless theory remain expressed informally in research papers and lack machine-readable representations. Unlike pure mathematics, where libraries such as Mathlib provide extensive formal coverage, wireless theory has no comparable repositories. Building domain-specific formal libraries is a prerequisite for enabling automated verification and reasoning across all three layers discussed in this Perspective.

\item Computational complexity of reasoning tasks.
As our case study reveals, the dominant failure modes in AI-assisted derivation are algebraic in nature rather than failures in problem understanding. This indicates that the principal bottleneck lies in reliably executing sustained symbolic manipulations, particularly as derivation depth increases. Strengthening symbolic backends and incorporating fine-grained, step-level verification mechanisms are among the most impactful targets for improving reasoning reliability.

\item Lack of domain-specific reasoning datasets.
Existing mathematical reasoning benchmarks focus on competition and undergraduate-level problems, and do not capture the multi-step, cross-domain derivations characteristic of wireless theory. Richer datasets encompassing full derivation chains with intermediate verification checkpoints are needed to train, fine-tune, and meaningfully evaluate reasoning systems on wireless mathematical tasks.
\end{itemize}

\section{Outlook}

Wireless communication theory contains a rich collection of mathematical reasoning problems that have long relied on manual derivation and domain expertise. Recent advances in AI reasoning systems, spanning formal theorem proving, neural-symbolic methods, and LLM-based derivation, provide an opportunity to transform how these problems are analyzed and solved.

We have outlined three layers through which AI-assisted reasoning may contribute to wireless theory: verifying established results, assisting in derivation, and discovering new theoretical foundations. These layers reflect a progressive roadmap from tool-augmented verification toward autonomous mathematical reasoning. Near-term progress is most tractable at the verification layer, where formalizing classical wireless results into machine-readable libraries would immediately benefit both proof tools and reasoning models. Extending this capability to derivation and discovery will require wireless-specific formal libraries, domain-adapted benchmarks, and reasoning systems capable of operating across the diverse mathematical domains that a single wireless derivation may span. As emerging architectures such as pinching antenna arrays and stacked intelligent metasurfaces give rise to new analytical frameworks, the demand for such capabilities will only grow. Beyond static analytical frameworks, AI-assisted reasoning will also need to accommodate the dynamic nature of wireless systems, where user mobility, carrier frequency, and propagation environments jointly shape the validity of any analytical result. Incorporating such dynamics into formal reasoning frameworks is essential for practical deployment. Addressing these challenges will require sustained collaboration between the AI reasoning and wireless communications communities, a convergence whose realization holds significant promise for the theoretical foundations of next-generation networks.